# Volume holograms with linear diffraction efficiency relation by (3+1)D printing

NIYAZI ULAS DINC,[1,2, *] CHRISTOPHE MOSER,[1] DEMETRI PSALTIS[2]

[1]*Laboratory of Applied Photonics Devices (LAPD), École Polytechnique Fédérale de Lausanne (EPFL), Lausanne, Switzerland*
[2]*Optics Laboratory (LO), École Polytechnique Fédérale de Lausanne (EPFL), Lausanne, Switzerland*

*[niyazi.dinc@epfl.ch](mailto:niyazi.dinc@epfl.ch)

**We demonstrate the fabrication of volume holograms using 2-Photon polymerization with dynamic control of light exposure. We refer to our method as (3+1)D printing. Volume holograms that are recorded by interfering reference and signal beams have a diffraction efficiency relation that is inversely proportional with the square of the number of superimposed holograms. By using (3+1)D printing for fabrication, the refractive index of each voxel is created independently and thus by, digitally filtering the undesired interference terms, the diffraction efficiency is now inversely proportional to the number of multiplexed gratings. We experimentally demonstrated this linear dependence by recording M=50 volume gratings. To the best of our knowledge, this is the first experimental demonstration of distributed volume holograms that overcome the 1/M² limit.**

The utilization of volume holograms has garnered significant attention in various applications such as optical interconnects [1], data storage [2, 3], optical correlators [4, 5] and mode multiplexing/demultiplexing [6]. The rationale behind harnessing 3-Dimensional (3D) volumes to store and process information is inherently intuitive: the introduction of an additional dimension offers an expanded storage capacity when compared to 2D optical layouts [7, 8]. However, a major challenge for volume holograms is the diffraction efficiency, defined as the fraction of the light power diffracted by the hologram to the incident power, which falls inversely with the square of the number of multiplexed holograms when the holograms are recorded with optical interference [9]:

$$\eta = \frac{(M\#)^2}{M^2} \quad (1)$$

where $\eta$ is diffraction efficiency, $M$ is the number of multiplexed holograms and ($M\#$) is the system metric to quantify the medium's storage capacity, which depends on the dynamic range of the refractive index (RI) and the thickness of the medium [10]. The dependence on the square of the number of holograms arises from the undesired terms that are included in the interference between the reference ($E_R$) and signal ($E_S$) beams during optical recording:

$$|E_R + E_S|^2 = E_R E_R^* + E_S E_S^* + E_R^* E_S + E_R E_S^* \quad (2)$$

The RI of the medium is modulated by the 3D intensity pattern given in Eq. (2). For the read-out, the reference beam illuminates the medium. The first term is usually a DC term since the reference beam is typically a plane wave. The second term contributes to noise by scattering the light in an undesired manner. The third term is the reconstruction term of the signal beam whereas the last term is the conjugate of the reconstruction term.

We can analyze the recording of multiple holograms by considering the summation of modulated sinusoids in 1D without loss of generality. We express the RI as:

$$\Delta n(x) = \sum_{m=1}^{M} A_m(x) \quad (3)$$

where $A_m$ is the index modulation due to a single recording, which can be expressed by a DC term and an AC term modulated by a data envelope ($f_m$) with oscillation frequency ($\omega_m$), resulting from the angle between the reference and the signal:

$$A_m = c_{DC} + c_{AC} \sin(\omega_m x + \varphi_m) \times f_m(x) \quad (4)$$

When we substitute Eq. (4) into Eq. (3), we find that the DC term grows linearly with M whereas the signal term only grows as the square root of M for large M:

$$\Delta n \propto M + \sqrt{M} \quad (5)$$

For large M, we can simply neglect $\sqrt{M}$. With this further simplification, Eq. (5) simply states that each modulated sinusoid has an amplitude proportional to $1/M$. Note that the diffraction efficiency for the intensity of each sinusoidal grating is proportional to the square of its share of the dynamic range of $\Delta n$, yielding the $1/M^2$ trend. If we are somehow able to equate $c_{DC}$ to zero to prevent the DC buildup, then Eq. (5) becomes:

$$\Delta n \propto \sqrt{M} \quad (6)$$

Eq. (6) implies that each modulated sinusoid would have an amplitude proportional to $1/\sqrt{M}$, consequently yielding a diffraction efficiency relation with $1/M$ trend. This phenomenon was verified previously by recording localized holograms in separate slices of a doubly doped lithium niobate crystal, which were locally photosensitized prior to holographic recording [11]. However, for fully distributed holograms, DC buildup is unavoidable with optical distributed means of recording. We have previously demonstrated voxel-by-voxel arbitrary RI writing by using 2-photon polymerization, which we refer to as (3+1)D printing [12]. Here, we use this method with a commercially available Nanoscribe 2-photon printer to demonstrate that it is achievable to obtain a linear diffraction efficiency relation for non-localized (distributed) volume holograms as well. First, we compute digitally the index distribution Δn(x, y, z) for volume holograms that consist of only the superimposed volume gratings without the first two terms of Eq. (2). Note that we need to keep the conjugate of the reconstruction term to obtain real-valued index distribution. Using the Beam Propagation Method (BPM), we simulated the wave propagation [13] and verified in simulation the linear trend as shown in Fig. 1 for 200 μm and 400 μm thick holograms in transmission geometry (see Supplementary Information Section 1 for more details). In Fig. 1, we clearly observe the expected linear dependence. For reference, we also plot $1/M$ and $1/M^2$ lines. Note that doubling the thickness of the material roughly doubles the M# as theoretically expected, which is obtained by a linear fit to the data points for M>10.

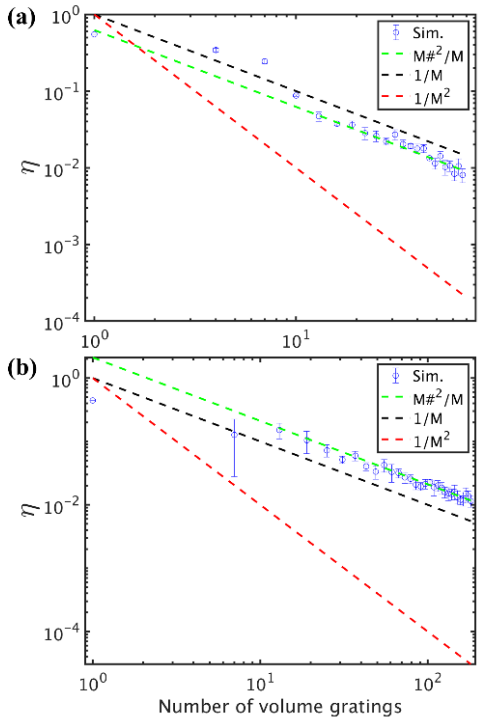

Fig. 1. Logarithmic plot for the simulated diffraction efficiency (y-axis) versus number of multiplexed volume gratings (x-axis) for (a) 200 μm thickness with ~70 volume gratings superimposed and (b) 400 μm thickness with ~200 volume gratings superimposed. Y-intercept of the plots correspond to M#$^2$.

Thanks to (3+1)D printing, we can directly fabricate these digitally calculated holograms. Instead of the IP-Dip resin, we used earlier [12]; we use IP-S, which has a smoother polymerization curve [14] enabling calibration-free fabrication by assuming near-linear dependence of RI vs. power. It is worthwhile to note that IP-Dip and IP-S resins have been used to fabricate various optical structures as reported in literature [15-20]. Here, we used 10 mm/s scanning speed as a fixed parameter and we varied the laser beam power to vary the polymerization exposure, which goes from 15% to 75% of the maximum power of the Nanoscribe system (when power scale is set to one), corresponding to an average power range from 3 mW to 15 mW. IP-S is photo-polymerized with a 780 nm femtosecond pulsed laser built-in in the Nanoscribe printer, focused by a 25x microscope objective. Hatching (lateral spacing between the centers of two voxels) and slicing (vertical spacing between the centers of two voxels) distances are 0.4 μm and 1 μm respectively. Hatching and slicing distances are important parameters, as they maintain a delicate balance between two key factors: structural stability and resolution on one hand, and printing speed on the other. Since we work in transmission geometry for holograms, the hatching distance is more critical as it sets the more demanding lateral resolution. A hatching distance of 0.4 μm is chosen as a good balance as we did not observe significant differences in diffraction orders when hatching is reduced (see Supplementary Information Section 2) suggesting that we become restricted by the polymerization process and diffusion.

We started the experimental characterization by printing, unslanted sinusoidal volume gratings and sweeping the grating period. In Fig 2(a-c) we provide Scanning Electron Microscope (SEM) images with three different periods. Since the voxel volume increases with the exposure, there is a surface relief pattern that forms and it is picked up by the electron microscope. This explains why the topography qualitatively shows the exposure map of the printed structure. In Fig. 2(d), we also provide a SEM image of a slanted volume grating, which clearly shows the slanted lines from the side. We decided to set Λ=2 μm as the minimum lateral period for this study since below this value, high index voxels merge together. This Λ value corresponds to a maximum incidence angle of approximately 13° for a wavelength λ=681 nm and average RI of the medium as 1.51. To find the dynamic range of RI with the given printing parameters, we printed unslanted volume gratings of different thicknesses of 110, 140, 170, and 200 μm with 6° Bragg angle and obtained the diffraction efficiency plot given in Fig. 2(e). To prevent additional diffraction from the air-polymer interface, an additional 5-μm thick homogeneous layer is printed on top. Using coupled wave theory [21, 22] we extracted the RI variance Δn= $1.7 \times 10^{-3}$ using Eq. (7).

$$\eta = \sin^2\left(\frac{\pi n_1 L}{\lambda \cos\theta}\right) \qquad (7)$$

where the amplitude of index variation is $n_1 = \Delta n/2$, $L$ is the thickness, and $\theta$ is the Bragg angle.

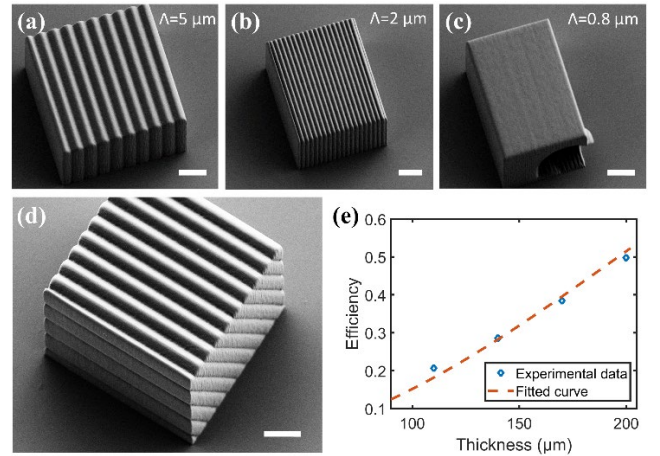

Fig. 2. (a-c) SEM images of unslanted sinusoidal volume gratings with periods 5 μm, 2 μm, and 0.8 μm respectively. (d) SEM image of a slanted volume grating. Scale bars measure 10 μm. (e) Diffraction efficiency of single

gratings of different thicknesses along with sin² curve fitting yielding the dynamic range of RI.

We printed the holograms as cubes with volume of $(200~\mu m)^3$ in accordance with the field of view and the working distance in the printing configuration. We can directly deduce the M# from the last data point in Fig 2(e), which corresponds to 200 μm thickness. Hence, the M# of our holograms is the square root of 0.5, resulting in roughly 0.7 for the given Bragg angle.

As well as RI, the voxel size varies with the optical power. Larger voxel size effectively decreases the resolution. For higher carrier frequencies, this inevitably results in a smoothening effect. Since the higher carrier frequencies are distorted, we see a frequency response in diffraction efficiency measurements. To probe this effect, we fabricated a hologram by multiplexing five gratings with equal strengths that were deigned to be Bragg-matched to the same signal beam (designed to be in the same direction) at five different reference beam angles. Since diffraction is a linear phenomenon, we built a reciprocal experimental setup where the illumination is such that it excites all the recorded gratings simultaneously. In this way, we can read-out all five gratings at the same time with a single beam illumination (see Supplementary Information Section 2). We schematically show this reciprocity in Fig. 3(a) using k-vectors of incident and diffracted beams and the gratings within the Ewald's sphere representation. In Fig. 3(b), we show the camera capture where we see all the five diffracted beams. We observe that the diffracted power decrease as the carrier frequency or the corresponding angle increases. We simulated the beam propagation in the calculated hologram to mimic the experiment. As theoretically expected, all the diffracted signals have equal strengths as shown on the cut line plot in Fig. 3(c). We hypothesize that the decrease of diffraction efficiency is caused by the smoothing of the sinusoidal grating during fabrication due to the voxel size dependence on optical power. To test this hypothesis, we model the smoothening by convolving the index distribution of each grating with a 2D Gaussian smoothing kernel on each slice of the calculated hologram. As we vary the standard deviation (σ) of the Gaussian kernel, we observed decreasing diffracted signal power as the angle increases mimicking our experimental observation. In Fig. 3(c), we provide the cutline plot of the simulated diffraction pattern with no Gaussian filtering, along with cutline plots obtained with filtered holograms for σ=1 and σ=2.1 and the experimentally recorded data points. We note that the smoothened hologram with σ=2.1 roughly corresponds to the decreasing power trend experimentally measured.

In order to perform diffraction efficiency measurements as a function of the number of recorded gratings independently of the frequency response, we printed three $(200~\mu m)^3$ cubes with 10, 30, and 50 multiplexed gratings and we measured the diffraction efficiency of the ones with the same spatial frequency. This is approximately 7.2° between the reference and the signal beams (see Supplementary Information Section 2). In Fig. 3(d), we show the experimentally measured diffraction efficiencies in log-log scale. We applied a linear curve fit and constrained y-intercept to be less than 0.5 as we know that the baseline efficiency should be smaller than the one we found using single grating with 6° Bragg angle because of the frequency response. In Fig. 3(d), we also provide the fitting curve, which yields a slope of -1.097 and y-intercept of 0.38 (M#=0.62). For reference, we also plot the lines with slopes -1 and -2, having the same M#.

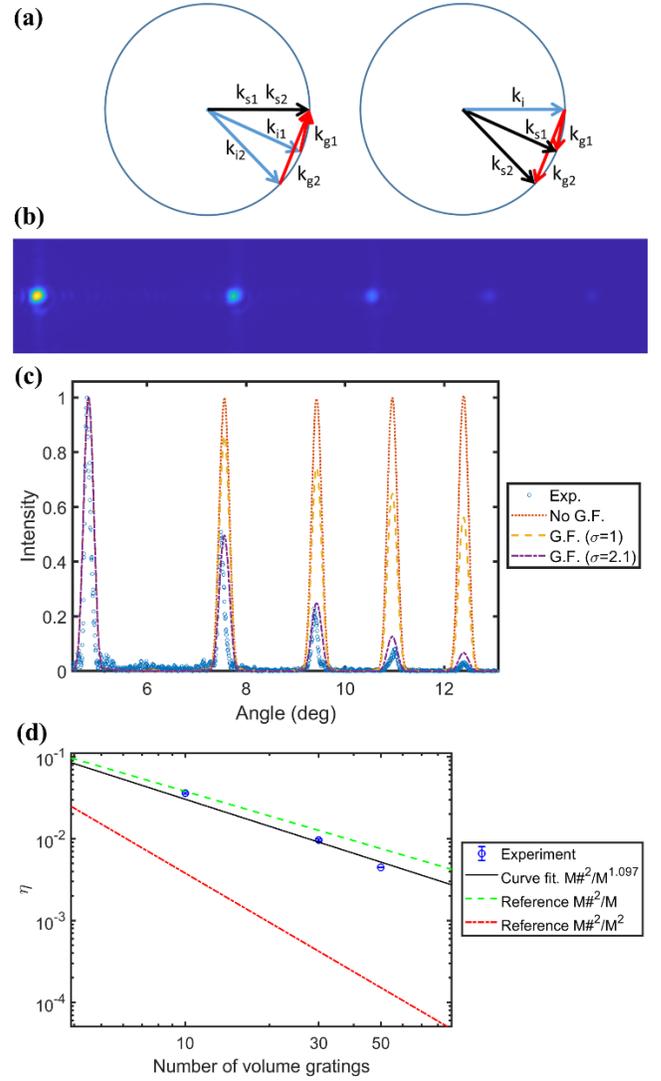

Fig. 3. (a) Ewald sphere representation, highlighting momentum matching for two different incidence beams in forward pass through a transmission hologram on the left. On the right, the reciprocal excitation of all the recorded gratings. (b) Camera capture of diffracted signals from the printed hologram. (c) Cut line plot showing the experimental data along with simulation data where G.F. stands for Gaussian-Filtered. (d) Experiment results showing near-linear diffraction efficiency trend along with reference lines of M#²/M and M#²/M².

Although (3+1)D printing paves way for digital optimization of the RI design, having a photoresist that has a frequency response because of the varying voxel size hampers the diffraction efficiency for higher carrier frequencies. We argue that this effect might also be responsible for the discrepancy of the experimental finding from the exact linear relation, although this argument requires more investigation as a future work. A photoresist whose RI varies without a significant change of voxel size is desired. However, we acknowledge that achieving such a chemistry is very challenging. Moreover, scaling the volume of the printed structures is crucial to reach higher storage capacities. Because of the limited working distance and field of view in photo polymerization, this can be achieved by employing stitching of different blocks. The risk of stitching is having a phase difference in different blocks, which would yield a distorted reconstruction. We have performed a preliminary study to show that stitching is feasible by observing SEM images showing slated grating periods in phase among different blocks (see Fig. 4). This is achieved when the printing is performed layer by layer for all the structure instead of block by block, which is consistent with our previous study [18].

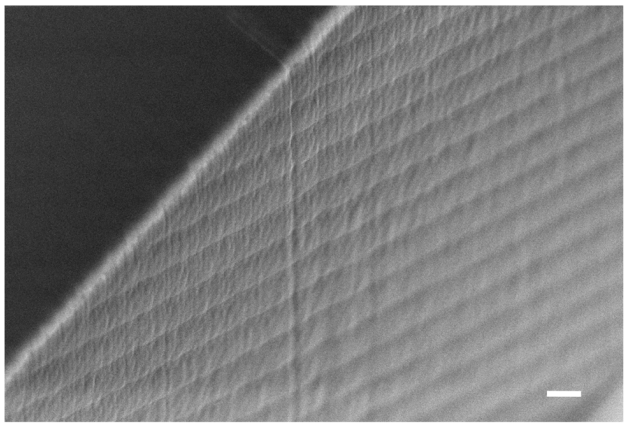

Fig. 4. SEM image of stitched blocks fabricated in adjacent field of views of the printing system. Block boundary is visible by the vertical line. The grating lines are visibly in-phase. The scale bar measures 2 μm.

Scaling is still challenging with respect to fabrication time since 2-photon-polymerization is a point-scanning technique. Another technical difficulty arises from the fact that we need to have all the voxels to be printed stored in the memory of the printing computer with assigned coordinates and dynamic laser power values, which becomes demanding for volumes greater than $(200~\mu m)^3$ for the 8 GB memory that is available in the facility. Nonetheless this issue can be solved by optimizing the printing protocols for (3+1)D printing.

Overall, this study reports on an important milestone for optical data storage, which is the experimental demonstration that the diffraction efficiency of M holograms is inversely related to M rather than $M^2$. This has been achieved by writing the computed holograms voxel by voxel using 2-photon printing.

**Disclosures.** The authors declare no conflicts of interest.

**Data availability.** Data underlying the results presented in this paper may be obtained from the authors upon reasonable request.

# Supplementary Information for "Volume holograms with linear diffraction efficiency relation by (3+1)D printing"

**NIYAZI ULAS DINC,**[1,2,*] **CHRISTOPHE MOSER,**[1] **DEMETRI PSALTIS**[2]

[1]*Laboratory of Applied Photonics Devices (LAPD), École Polytechnique Fédérale de Lausanne (EPFL), Lausanne, Switzerland*
[2]*Optics Laboratory (LO), École Polytechnique Fédérale de Lausanne (EPFL), Lausanne, Switzerland*
*[niyazi.dinc@epfl.ch](niyazi.dinc@epfl.ch)*

**Section 1: Numerical investigations**

All the numerical studies are performed on MATLAB using Beam Propagation Method. To compute the index distribution of volume holograms, we simply generate the reference beam and signal beam in the input aperture and let them propagate within the volume of interest. Then we record the 3D field distribution of both beams and compute the filtered interference by simply calculating:

$$E_{filtered,i}(x,y,z) = E_{R,i}^*(x,y,z)E_{S,i}(x,y,z) + E_{R,i}(x,y,z)E_{S,i}^*(x,y,z)$$

Where $i$ denotes the individual hologram. This way we get a real valued $E_{filtered,i}(x,y,z)$ without undesired DC and noise terms. Then, we simply add up all the computed filtered interferences and scale them according to the dynamic range of refractive index available. Since the reference and signal fields are generated with equal strengths, each hologram approximately shares an equal portion of the dynamic range. To compute the angle of the reference beams, we first compute the available carrier frequencies using Bragg selectivity curves computed for the given volume, background index and the wavelength. The carrier frequencies are computed iteratively by placing the peak of the subsequent one on the first zero crossing of Bragg selectivity curve of the preceding one. This is necessary as we work on small angles for a small volume that yields significant changes in the Bragg selectivity curves. For 200-µm thickness, we have computed 68 gratings to be stored by angular and peristrophic multiplexing. Among which we sampled these holograms with a step of 3 (M=1, 4, 7 …) and performed beam propagation on each hologram by scanning input angle and recorded the maxima intensities of the diffracted orders to calculate the mean and standard deviation of the diffraction efficiency. For 400-µm thickness, we have computed 183 gratings to be stored by angular and peristrophic multiplexing. Among which we sampled these holograms with a step of 6 (M=1, 7, 13 …). We used a RI dynamic range of $2 \times 10^{-3}$, wavelength λ=681 nm and average RI of the medium as 1.51. The angular sampling is 0.025 degrees for 200-µm thickness and 0.015 degrees for 400-µm thickness.

The 1/M trend assumes statistically independent phase relationship among recorded holograms for a large number of M so that the summation scales with $\sqrt{M}$. If this is not the case, the summation will simply generate beat frequencies in which all the peaks will add up and scale with $M$ instead of $\sqrt{M}$, which would yield 1/M² diffraction efficiency relation. We introduced a random phase bias to each recording to prevent this phenomenon. Hence, we were able to see the 1/M trend in the simulations even though there is a small oscillation in the mean and standard deviation values. For larger M than we used, we expect a more stable linear trend. Moreover, since the index distribution is calculated digitally, one can minimize or eliminate the generation of beat frequencies using various methods. A naïve and straightforward approach is sweeping initialized random phases or apply clipping without significantly distorting stored holograms. In our numerical analysis, we did not apply any further restrictions or iterations since the resulting trend was already satisfactory.

## Section 2: Experiments and characterizations

To remove the remaining monomer after the printing, development in PGMEA and IPA is performed for 10 minutes and 4 minutes respectively. In Fig. S1 we provide some snapshots of the excited orders for different holograms we studied during the preliminary phase where we see the desired orders and not the conjugates of those, which is a sanity check showing we indeed multiplexed volume gratings. The number of multiplexed gratings are indicated on the top left corner of each image.

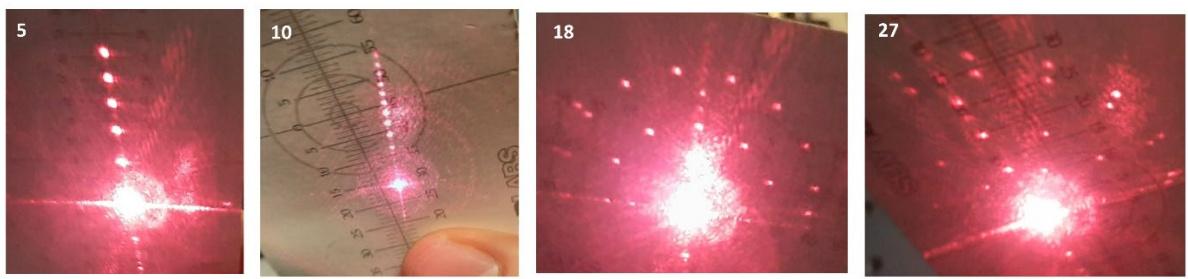

Supplementary Figure 1: Various snapshots during the preliminary study phase: the diffracted orders obtained when the hologram is flipped and illuminated.

For quantitative measurements, we have characterized the holograms by using the experimental setup whose schematic is given in Fig. S2.

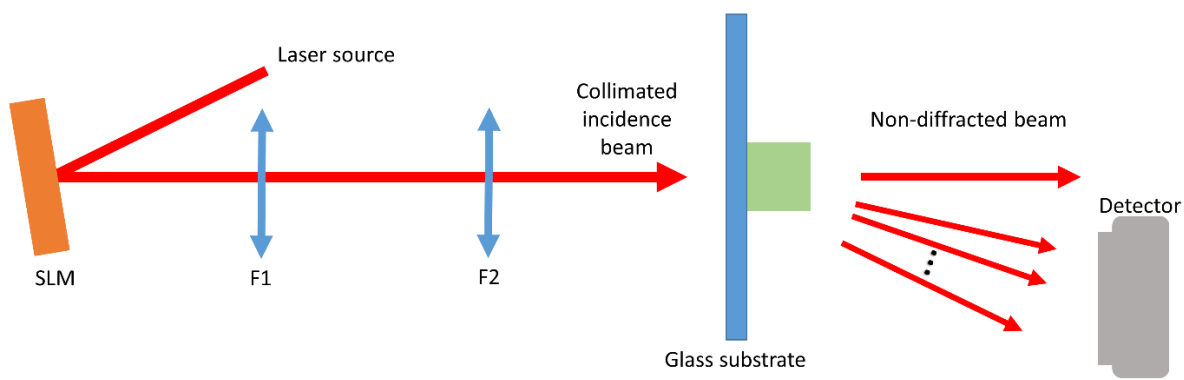

Supplementary Figure 2: Schematic of the experimental setup for diffraction efficiency measurements.

We also provide two photographs of the experimental setup in Fig. S3. We used the 4F system indicated to bring the holograms in the center of the illumination beam. Then we flip the front lens of the 4F imaging system and put a camera to realize the setup depicted in Fig. S2. We used the SLM to change the angle of the illumination beam to match the tilt angle of the cover slip, which serves as the substrate of the holograms printed on of it. By doing so, we illuminate the holograms with the perpendicular incidence angle, which maximizes the power in the diffracted orders.

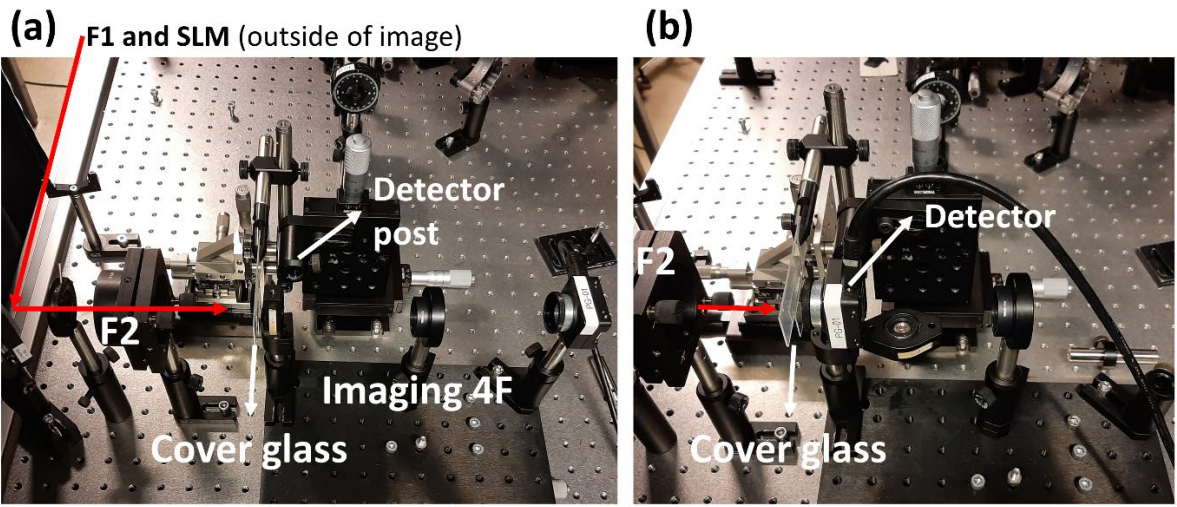

Supplementary Figure 3: Photographs of the experimental setup for diffraction efficiency measurements. (a) 4F imaging condition where there is a 4F system after the cover glass to align the samples with respect to illumination beam. (b) Configuration for capturing diffraction orders where we place the detector after the cover glass (notice that the front mirror of the imaging 4F is flipped to make room for the detector).

As mentioned in the main text we fabricated a volume hologram that consists five volume gratings. In order to probe if the chosen hatching distance parameter is appropriate, meaning that it provides sufficient sampling, we printed this hologram with different hatching parameters (i.e. different XY sampling of the printer beam trajectory). We provide the results in Fig. S3 showing that there is no significant change in the plots, especially for high angles or large carrier frequencies. This is to say that we are restricted by polymerization chemistry rather than the sampling of the trajectory of the printing beam. Note that to model the smoothening (as in the plots shown in Fig. 3 of the main text) we used "imgaussfilt" function of MATLAB where the standard deviation is set by the "sigma" argument of this function.

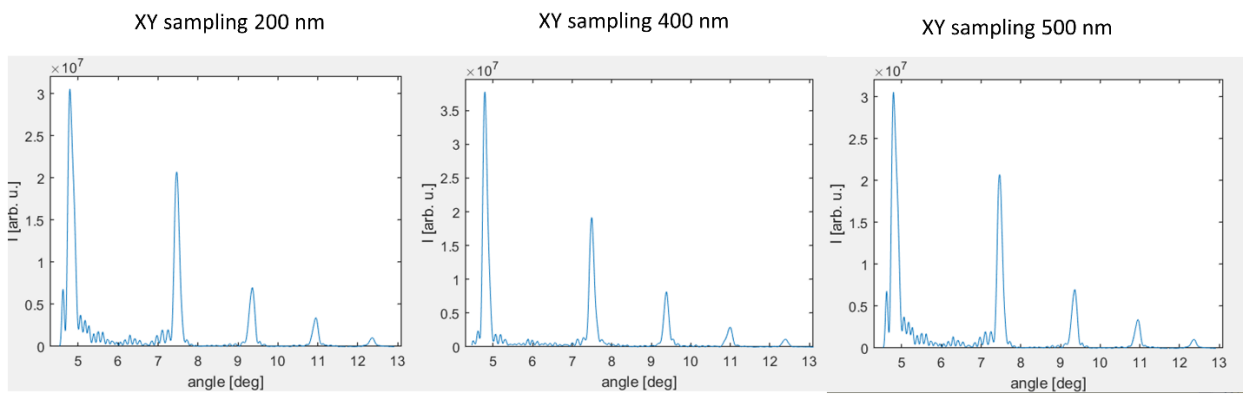

Supplementary Figure 4: Cut line plot showing the experimental data of diffracted orders corresponding to five volume gratings superimposed. From left to right, the hatching distances (or XY sampling of the trajectory of the printing beam) of the holograms are 200 nm, 400 nm and 500 nm respectively.

In Fig. S5, we show the Fourier transform of the refractive index distribution of middle z-slice of each hologram generated by multiplexing 10, 30, and 50 volume gratings respectively from left to right. Hence the axes represent spatial frequencies in X and Y. Since we just take a slice in z from each hologram, we see the orders and the conjugate terms. Recall that the 3D nature prevents the excitation of conjugate terms. We also indicate the orders, which correspond to approximately 7.2° in polar direction, used to calculate diffraction efficiency by red arrows.

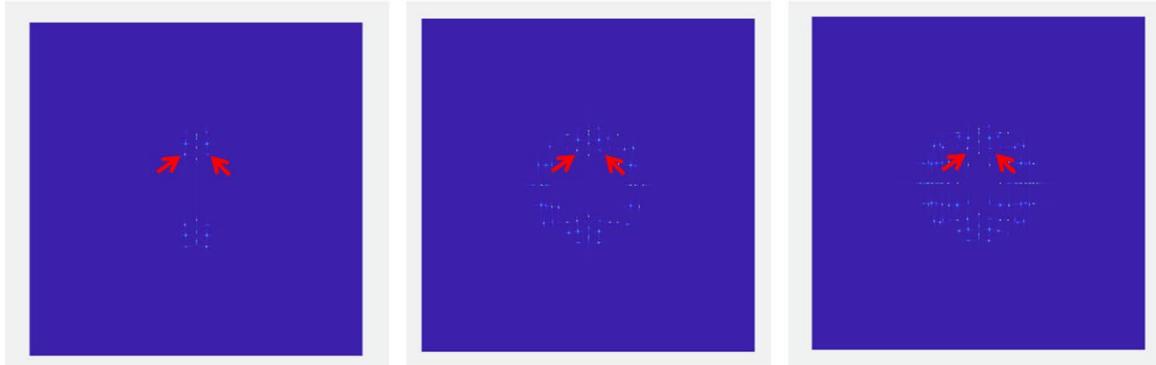

Supplementary Figure 5: Fourier transforms of the refractive index distribution of mid-z plane of each hologram generated by multiplexing 10, 30, and 50 volume gratings respectively from left to right. Red arrows indicate the orders with the same magnitude of spatial frequency used to compare the diffraction efficiencies of different holograms.